\documentclass[prd,aps,showpacs,nofootinbib,preprint,eqsecnum]{revtex4}
\usepackage{graphicx,color,amsmath,amsxtra}
\usepackage{amssymb}
\usepackage[english]{babel}
\usepackage{amsfonts}
\allowdisplaybreaks[4]

\begin{document}

\title{Scalar field dark energy with a minimal coupling in a spherically symmetric background}

\author{Jiro Matsumoto\footnote{E-mail address: jmatsumoto@kpfu.ru}
}
\affiliation{Institute of Physics, Kazan Federal University, Kremlevskaya Street 18,
Kazan 420008, Russia}
	
\begin{abstract}
Dark energy models and modified gravity theories have been actively studied and 
the behaviors in the solar system have been also carefully investigated in a part of the models. 
However, the isotropic solutions of the field equations 
in the simple models of dark energy, e.g. quintessence model without matter coupling, 
have not been well investigated. 
One of the reason would be the nonlinearity of the field equations. 
In this paper, a method to 
evaluate the solution of the field equations is constructed, 
and it is shown that there is a model that can easily pass the 
solar system tests, whereas, there is also a model that is constrained from the solar system tests. 
\end{abstract}
	
	
\maketitle

\section{Introduction \label{sec1}}
The accelerated expansion of the Universe was discovered in the late 1990s from the 
observations of type Ia supernovae \cite{Riess:1998cb,Perlmutter:1998np}. 
Many hypotheses about the cause of the accelerated expansion have been proposed, 
however, it is still unclear what is the real cause of it. 
The representative way to explain the accelerated expansion of the Universe are the following; 
introducing the cosmological constant $\Lambda$, introducing a dynamical scalar field 
instead of $\Lambda$, modifying the geometry part of the Einstein equations.   
They are called dark energy models or modified gravity theories. 
We will treat $k$-essence model \cite{ArmendarizPicon:1999rj,Garriga:1999vw,ArmendarizPicon:2000ah}, 
which is a scalar field model of dark energy and contains quintessence model 
\cite{Peebles:1987ek,Ratra:1987rm,Chiba:1997ej,Zlatev:1998tr} as a special case, in this paper.  
$k$-essence model has only minimal coupling with gravity, the potential term of the field, and the kinetic terms of the field. 

Scalar field models of dark energy are defined not by the field equations but by their action. 
Therefore, not only homogeneous, isotropic, and expanding solutions but also static spherical solutions 
are affected by the scalar field. 
The influences for the static spherical solution from dark energy 
have been considered in the models that have a nonminimal coupling with 
the usual matter \cite{Khoury:2003rn} 
or have a nonminimal coupling with gravity \cite{Will:2014kxa}, 
however, they have not been enough investigated in $k$-essence model without matter coupling. 
One of the reason may come from an assumption that 
there are no serious influences to gravity from a minimal couping. 
In fact, the gravitational effects from the cosmological constant in the solar system certainly exist, however, 
they are too small to be observed \cite{Kagramanova:2006ax}. While, the dynamical scalar field under 
the Friedmann-Lemeitre-Robertson-Walker (FLRW) space-time 
varies depending on the radius $r$ in a spherically symmetric background. 
Therefore, we should carefully take into account the influences from the scalar field because 
they can be large enough to be observed in the solar system even if they are same order as those from the cosmological constant 
at the horizon scale of the Universe. 
The other reason would be caused from the nonlinearity of the field equation. 
The analytic solutions for the nonlinear differential equation do not generally exist. 
The reason why the investigations in more complex theories are possible is that 
many assumptions are applied. 
It is important to inspect whether or not the assumptions are valid in a simpler model. 
In this paper, we will consider the behavior of the scalar field in 
a static and spherical space-time in $k$-essence model without matter coupling. 

The contents of the paper are as follows. We 
deduce the general equations in a static and spherical background, and construct a general formalism to evaluate the 
solution in Sec.~\ref{sec2}. 
The solutions of the equations in quintessence model are investigated in Sec.~\ref{sec3}, and 
those in $k$-essence model are considered in Sec.~\ref{sec4}. Conclusions are in Sec.~\ref{sec5}. 
The units of $k_\mathrm{B} = c = \hbar = 1$ are used and 
gravitational constant $8 \pi G$ is denoted by
${\kappa}^2 \equiv 8\pi/{M_{\mathrm{Pl}}}^2$ 
with the Planck mass of $M_{\mathrm{Pl}} = G^{-1/2} = 1.2 \times 10^{19}$GeV in this paper. 
\section{Construction of a general formalism for evaluating the solution \label{sec2}} 
We consider the following action of $k$-essence model: 
\begin{equation}
S = \int d^4x \left [ \frac{R}{2 \kappa ^2} - K(\phi , X) \right ] + S_\mathrm{matter}, 
\label{10}
\end{equation}
where $\phi$ is a scalar field, $X$ is a kinetic term of the scalar field 
$X \equiv - \partial _ \mu \phi \partial ^\mu \phi /2$ , and $K$ is an arbitrary function of $\phi$ and $X$. 
$S_\mathrm{matter}$ is the action of the usual matter. 
The equations given from the principle of least action are given as follows: 
\begin{align}
\frac{1}{\kappa ^2} G_{\mu} ^\nu 
= -\delta _\mu ^\nu K(\phi , X) - K_{, X} (\phi ,X) \partial _\mu \phi \partial ^\nu \phi + T_\mu ^{\nu \; \mathrm{(matter)}}, 
\label{20} \\ 
K_{, \phi} (\phi , X) + \nabla _\mu (K_{,X}(\phi ,X) \partial ^\mu \phi) =0, 
\label{30}
\end{align}
where the subscript $_{, A}$ means derivative with respect to $A$. 
If we assume a static metric with a spherical symmetry, 
\begin{equation}
ds^2 = - \mathrm{e} ^{2 \Phi (r)} dt^2 + \mathrm{e} ^{2 \lambda (r)} dr^2 + r^2 (d \theta ^2 + \sin ^2 \theta d \varphi ^2), 
\label{40}
\end{equation}
we obtain 
\begin{align}
\frac{1}{\kappa ^2} \left ( \frac{1}{r^2} - \frac{2 \lambda '}{r} \right ) \mathrm{e} ^{-2 \lambda} - \frac{1}{\kappa ^2 r^2} = 
- K+ \mathrm{e} ^{-2 \Phi} \dot \phi ^2 K_{,X} - \rho _\mathrm{m} 
\label{50}, \\
0= - \mathrm{e}^{-2 \lambda} \dot \phi \phi ' K_{,X}
\label{60}, \\
\frac{1}{\kappa ^2} \left ( \frac{1}{r^2} + \frac{2 \Phi '}{r} \right ) \mathrm{e} ^{-2 \lambda} - \frac{1}{\kappa ^2 r^2} = 
- K -\mathrm{e} ^{-2 \lambda} \phi ^{\prime 2} K_{,X} + p_\mathrm{m}
\label{70}, \\
0= \frac{1}{r^2} \phi ' \partial _ \theta \phi K_{,X}
\label{80}, \\
\frac{1}{\kappa ^2} \left ( \Phi '' + \Phi ^{\prime 2} - \lambda ' \Phi ' + \frac{\Phi '}{r} - \frac{\lambda '}{r} \right ) \mathrm{e} ^{-2 \lambda} = 
-K - \frac{1}{r^2} ( \partial _\theta \phi )^2 K_{, X} + p_\mathrm{m} 
\label{90}, \\
0= \frac{1}{r^2 \sin ^2 \theta} \phi ' \partial _ \varphi \phi K_{,X}
\label{100},
\end{align}
where dots and primes mean time derivatives and the derivatives with respect to $r$, respectively. 
$\rho _\mathrm{m}$ and $p_\mathrm{m}$ represent 
the energy density and the pressure of the usual matter, respectively. 
If $\phi ' \neq 0$ and $K_{, X} \neq 0$ are assumed, we obtain $\dot \phi = \partial _\theta \phi = \partial _\varphi \phi = 0$ 
from Eqs.~(\ref{60}), (\ref{80}), and (\ref{100}). 
By using these conditions, we can obtain the following equation from the field equation (\ref{30}): 
\begin{align}
(K_{,X}- \mathrm{e}^{-2 \lambda} K_{,XX} \phi ^{\prime 2}) \phi '' 
&+ \left (K_{, X \phi} \phi ' + \frac{2}{r}K_{,X} \right ) \phi ' + \mathrm{e}^{2 \lambda} K_{, \phi} \nonumber \\
-& (K_{,X} - \mathrm{e} ^{-2 \lambda} K_{,XX} \phi ^{\prime 2}) \phi ' \lambda ' + K_{,X} \phi ' \Phi ' =0. 
\label{110}
\end{align}
Whereas, the equations of continuity $\nabla  _\mu T_\nu ^{\mu \; \mathrm{(matter)}}=0$ give 
\begin{align}
\dot \rho _\mathrm{m} =0, \label{120} \\
p _\mathrm{m}' + \Phi ' (\rho _\mathrm{m} + p_\mathrm{m}) =0. \label{130}
\end{align}
If we assume $\vert \Phi \vert , \vert \lambda \vert \ll 1$ and $\rho _\mathrm{m} = p_\mathrm{m} =0$, 
we can express $\Phi (r)$ and $\lambda (r)$ as $\Phi (r) = r_0/r +\delta \Phi (r)$, where $r_0$ is a constant, 
and $\lambda (r)= - \Phi (r)+ \delta \lambda (r)$, 
because Eqs.~(\ref{50}) and (\ref{70}) can be rewritten by 
\begin{align}
\lambda ' + \frac{1}{r} \lambda = \frac{\kappa ^2 r}{2} K, 
\label{150}\\
\Phi ' - \frac{1}{r} \lambda = - \frac{\kappa ^2 r}{2} \left [ K+(1-2 \lambda)\phi ^{\prime 2}K_{,X} \right ], 
\label{160}
\end{align} 
at the leading order in $\Phi$ and $\lambda$. 
If the right-hand-side of Eqs.~(\ref{150}) and (\ref{160}) vanish, then 
$\Phi (r) = r_0/r$ and $\lambda = - \Phi$ are the solution set of them. 
Therefore, $\delta \Phi$ and $\delta \lambda$ express the deviations from the 
vacuum solution. 
If $\vert r_0/r \vert$, $\vert \delta \Phi \vert$, $\vert r \delta \Phi ' \vert$, $\vert \delta \lambda \vert$, and 
$\vert r \delta \lambda ' \vert$ are 
much less than $1$, 
then $\kappa ^2 r K$ and $\kappa ^2 r \phi ^{\prime 2}K_{,X}$ in Eqs.~(\ref{150}) and 
(\ref{160}) are also much less than $1$, thus, the term $\lambda \kappa ^2 r \phi ^{\prime 2}K_{,X}$ in Eq.~(\ref{160})
can be ignored. 
Then, $\delta \lambda '(r)$ is expressed as 
\begin{equation}
\delta \lambda ' (r) = -\frac{\kappa ^2 r}{2} \phi ^{\prime 2}(r) K_{,X}(\phi (r),X(r)). 
\label{170}
\end{equation}
The expression of $\delta \Phi$ is given by solving Eq.~(\ref{160}) as 
\begin{equation}
\delta \Phi (r)= \frac{1}{r} \int ^r dl \left \{ \delta \lambda (l)- \frac{\kappa ^2 l^2}{2} 
\left [ K(\phi (l),X(l)) +\phi ^{\prime 2} K_{,X}(\phi (l),X(l)) \right ] \right \}
+\frac{r_1}{r}, 
\label{180}
\end{equation}
where $r_1$ is an integration constant. Eqs.~(\ref{170}) and (\ref{180}) show that 
the metric functions $\delta \Phi (r)$ and $\delta \lambda (r)$ depend on the 
scalar field in a nontrivial way. 
Not only the values of $\phi (r)$ and $\phi '(r)$ but also the form of the potential 
explicitly affect the metric functions. 
Whereas, Eq.~(\ref{110}) in the case $\rho _\mathrm{m} = p_\mathrm{m} =0$ and $\vert \Phi \vert , \vert \lambda \vert \ll 1$ 
is expressed by 
\begin{align}
\bigg [ K_{,X} -(1-2 \lambda) K_{,XX}\phi ^{\prime 2} \bigg ] \phi ''  
+ \bigg [ K_{,X} \frac{2}{r} +K_{,X \phi} \phi ' - (K_{,X}-K_{,XX}\phi ^{\prime 2})\lambda ' 
&+ K_{,X} \Phi ' \bigg ] \phi ' \nonumber \\
+(1+&2 \lambda) K_{, \phi} =0. 
\label{190}
\end{align}
If we only take care of the leading terms in Eq.~(\ref{190}), we have 
\begin{equation}
(K_{,X}-K_{,XX}\phi ^{\prime 2}) \phi '' + \left ( K_{,X} \frac{2}{r} +K_{, X \phi}\phi ' \right ) \phi ' + K_{, \phi}=0. 
\label{200}
\end{equation}
Equation (\ref{200}) is equivalent to the field equation in Minkowski space-time. 
The procedure to investigate the behavior of the scalar field in the vacuum is the following; 
first, to solve Eq.~(\ref{200}), second, substituting the solution of Eq.~(\ref{200}) into 
Eqs.~(\ref{170}) and (\ref{180}), third, checking whether the approximations 
$\vert \delta \lambda \vert , \vert r \delta \lambda ' \vert  , \vert \delta \Phi \vert , \vert r \delta \Phi ' \vert \ll 1$ 
are valid. 
In general, the conditions $\vert \delta \lambda \vert , \vert r \delta \lambda ' \vert  , \vert \delta \Phi \vert , \vert r \delta \Phi ' \vert \ll 1$ 
give a constraint for $r$, which is a domain of definition of Eq.~(\ref{200}).  
The condition $\vert r_0/r \vert \ll 1$ is always assumed because it is only $\sim 10^{-6}$ on the surface of the Sun. 
The behavior of the solution in the region that Eq.~(\ref{200}) is not valid is clarified by numerical calculations if 
we take the solution of Eq.~(\ref{200}) as a boundary condition. In the following sections, 
we will apply this procedure for the concrete examples. 
\section{Quintessence \label{sec3}}
In the case of quintessence model, the function $K(\phi, X)$ is written by
\begin{equation}
K(\phi, X) = -X +V(\phi) = \frac{1}{2} \mathrm{e}^{- 2 \lambda} \phi ^{\prime 2} + V(\phi), 
\label{q10}
\end{equation}
where $V$ is an arbitrary function of $\phi$. 
Then, Eq.~(\ref{200}) can be simplified as
\begin{equation}
\phi '' + \frac{2}{r}  \phi '-  V_{, \phi} =0. 
\label{q25}
\end{equation}
In general, it is difficult to solve Eq.~(\ref{q25}) because it is a nonlinear differential equation. 
The cases $V(\phi)=0 $\cite{Bergmann:1957zza} and $V(\phi)=m^2 \phi ^2 $\cite{Momeni:2009qw} have been already investigated. 
\subsection{The case $w=const.$}
The case that the equation of state parameter $w=p/ \rho$ is constant is often discussed in cosmology. 
A fluid with constant $p / \rho$, where $\rho$ is the energy density of the fluid and 
$p$ is a direction-averaged pressure, in a spherically symmetric space-time is discussed 
in \cite{Chernin:2001nu,GonzalezDiaz:2001ce,Kiselev:2002dx}, 
however, it is different from quintessence model with a constant equation of state parameter in cosmology, 
which is defined in 
the Friedmann-Lemeitre-Robertson-Walker (FLRW) space-time. 
In the flat FLRW background, $ds^2=-dt^2+a(t)\Sigma _i dx_i^2$, 
we have the following equations, 
\begin{align}
\frac{3 H^2}{\kappa ^2} = \frac{1}{2} \dot \phi ^2 + V + \rho _\mathrm{m}, \label{q30} \\
-\frac{1}{\kappa ^2}(2 \dot H + 3H^2) = \frac{1}{2} \dot \phi ^2 - V + p_\mathrm{m}, \label{q40}
\end{align} 
and 
\begin{equation}
\ddot \phi + 3 H \dot \phi + V_{, \phi} =0. \label{q45}
\end{equation}
Here, Eqs.~(\ref{q30}) and (\ref{q40}) come from the Einstein equations and Eq.~(\ref{q45}) 
is the equation of motion for the scalar field. 
If we impose $(\dot \phi ^2/2 -V)/(\dot \phi ^2/2 +V)=w=const.$, where $-1< w <+1$, we have 
\begin{equation}
V_{, \phi} = \frac{1-w}{1+w} \ddot \phi \label{q50}. 
\end{equation}
Substituting Eqs.~(\ref{q50}) into (\ref{q45}) yields 
\begin{equation}
\dot \phi = m^2 a^{-3(1+w)/2}, \label{q60}
\end{equation}
where $m^2$ is an integration constant. Therefore, we have 
\begin{equation}
V= \frac{1-w}{2(1+w)}m^4 a^{-3(1+w)}. 
\label{q70}
\end{equation}
Whereas, in the case of $\rho _\mathrm{m}=p_\mathrm{m}=0$, 
we can obtain 
\begin{equation}
a(t) = \left [ \frac{\kappa m^2 \sqrt{3(1+w)}}{2} \right ]^{\frac{2}{3(1+w)}} t^\frac{2}{3(1+w)} 
\label{q80}
\end{equation}
from Eq.~(\ref{q30}). Integration constant in Eq.~(\ref{q80}) is set by assuming $a(0)=0$. 
Substituting Eq.~(\ref{q80}) into (\ref{q60}) gives 
\begin{equation}
\dot \phi = \frac{2}{\kappa \sqrt{3(1+w)}t}, 
\label{q90}
\end{equation}
subsequently, we obtain
\begin{equation}
t = t_1 \exp \left [ \frac{\kappa \sqrt{3(1+w)}}{2} \phi \right ], 
\label{q100}
\end{equation}
where $t_1$ is a constant. 
Therefore, we finally obtain 
\begin{equation}
V(\phi) = \frac{2(1-w)}{3(1+w)^2} \frac{1}{t_1 ^2 \kappa ^2} \exp \left [ -\kappa \sqrt{3(1+w)}\phi \right ]. 
\label{q110}
\end{equation}
Equation (\ref{q110}) shows that 
the quintessence model, which yields a constant equation of state parameter, have 
an exponential potential.  
However, it is necessary to be careful that we assumed $\rho _\mathrm{m}=p_\mathrm{m}=0$, 
which is the approximation of dark energy dominance, 
in the calculations. 
The explicit form of the potential in the case $\rho _\mathrm{m}\neq 0$ is 
given in \cite{Sahni:1999gb}. 

In the following, we will investigate the behavior of the quintessence model 
with an exponential potential in a spherically symmetric background. 
Here, we use the following notation of the potential function $V(\phi)$: 
\begin{equation}
V(\phi) = M^4 \mathrm{e} ^{- \frac{\phi}{\phi _0}}. 
\label{q120}
\end{equation}
Then $M^4 \sim M_\mathrm{pl}^2 H_0 ^2$, where $H_0$ is the Hubble constant, and $\phi _0 \sim M_\mathrm{pl}$ 
should hold to realize the current expansion of the Universe. 
From the observations of supernovae, CMB, and BAO, $\phi _0$ is constrained to $2 M_\mathrm{pl} \lesssim \phi _0$ 
at $2 \sigma$ level \cite{Chiba:2012cb}. 
The field equation (\ref{q25}) is written by
\begin{equation}
\phi '' + \frac{2}{r} \phi ' + \frac{1}{\phi _0} M^4 \mathrm{e} ^{- \frac{\phi}{\phi _0}} =0. 
\label{q130}
\end{equation} 
If we assume $\vert \phi \vert \ll \phi _0$, 
we obtain 
\begin{equation}
\phi '' + \frac{2}{r} \phi ' + \frac{1}{\phi _0} M^4  \simeq 0. 
\label{q140}
\end{equation}
The validity of $\vert \phi \vert \ll \phi _0$ will be later considered. 
The solution of Eq.~(\ref{q140}) is obtained as 
\begin{equation}
\phi (r) = - \frac{M^4}{6 \phi _0} r^2 + \frac{c_1}{r} + m_1, 
\label{q150}
\end{equation}
where $c_1$ and $m_1$ are arbitrary constants. 
The behavior of $\phi ' $ varies depending on the length $r$ as seen from the right-hand side of Eq.~(\ref{q150}). 
The length $r_c$ that the former two terms in Eq.~(\ref{q150}) become same order is approximately given by 
$r_c \sim [c_1/ (M_\mathrm{pl}H_0^2)]^{1/3}$. 
The condition $\vert \phi \vert \ll \phi _0$ is now rewritten as $\vert c_1 \vert/M_\mathrm{pl} \ll r \ll H_0^{-1}$ if 
$m_1$ is much less than the Planck mass. 
The influence for the metric function $\lambda$ from the scalar field is evaluated by 
substituting Eq.~(\ref{q150}) into Eq.~(\ref{170}) as 
\begin{equation}
\delta \lambda '(r) = \frac{\kappa ^2 r}{2} \left ( \frac{M^8}{9 \phi _0 ^2}r^2 +\frac{2c_1 M^4}{3\phi _0 r} +\frac{c_1^2}{r^4} \right ). 
\label{q151}
\end{equation}
Integration of Eq.~(\ref{q151}) with respect to $r$ gives 
\begin{equation}
\delta \lambda (r) = \frac{\kappa ^2 M^8}{72 \phi _0 ^2}r^4 +\frac{c_1 \kappa ^2 M^4}{3\phi _0 }r 
+c_2 -\frac{c_1^2 \kappa ^2}{4 r^2}, 
\label{q152}
\end{equation}
where $c_2$ is an integration constant. Then, the conditions 
$\vert \delta \lambda \vert , \vert r \delta \lambda ' \vert \ll 1$ yield 
$r \ll H_0^{-1}$, $\vert c_2 \vert \ll 1$, and $\vert c_1/M_\mathrm{pl} \vert \ll r$. 
Whereas, substituting Eqs.~(\ref{q150}) and (\ref{q152}) into Eq.~(\ref{180}) 
gives 
\begin{equation}
\delta \Phi (r) \simeq \frac{\kappa ^2 M^8}{120 \phi _0^2}r^4 - \frac{\kappa ^2 M^4}{6}r^2 
+ \frac{c_1 \kappa ^2 M^4}{4 \phi_0}r +c_2 + \frac{r_1}{r}, 
\label{q153}
\end{equation}
where $\mathrm{e}^{-\phi / \phi _0}\simeq 1$ was applied and $r_1$ is an arbitrary constant. 
The condition $\vert \delta \Phi \vert \ll 1$ is, then, translated to 
$r \ll H_0^{-1}$, $\vert c_1H_0/M_\mathrm{pl} \cdot H_0 r \vert \ll 1$, $\vert c_2 \vert \ll 1$, and $\vert r_1 \vert \ll r$. 
Therefore, the solution (\ref{q150}) is valid in the region $\vert c_1/M_\mathrm{pl} \vert , \vert r_1 \vert \ll r \ll H_0^{-1}$ 
when the arbitrary constant $c_2$ is small enough. 
The values of the first terms in $\delta \lambda$ and $\delta \Phi$ are $\sim 10^{-72}$ for $r=10^9$ m, and the value of 
the second term in $\delta \Phi$ is $\sim 10^{-36}$ for $r=10^9$ m. Thus, this model can easily pass the solar system tests if 
we assign small values for the arbitrary constants $c_1, c_2$ and $r_1/GM_{\odot}$. 

\begin{figure}
\begin{minipage}[t]{0.5\columnwidth}
\begin{center}
\includegraphics[clip, width=0.97\columnwidth]{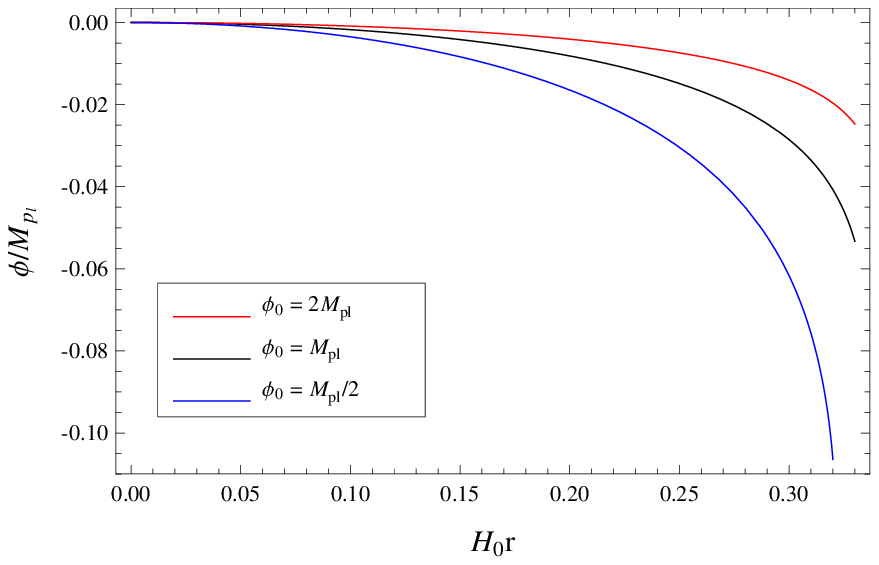}
\end{center}
\end{minipage}%
\begin{minipage}[t]{0.5\columnwidth}
\begin{center}
\includegraphics[clip, width=0.97\columnwidth]{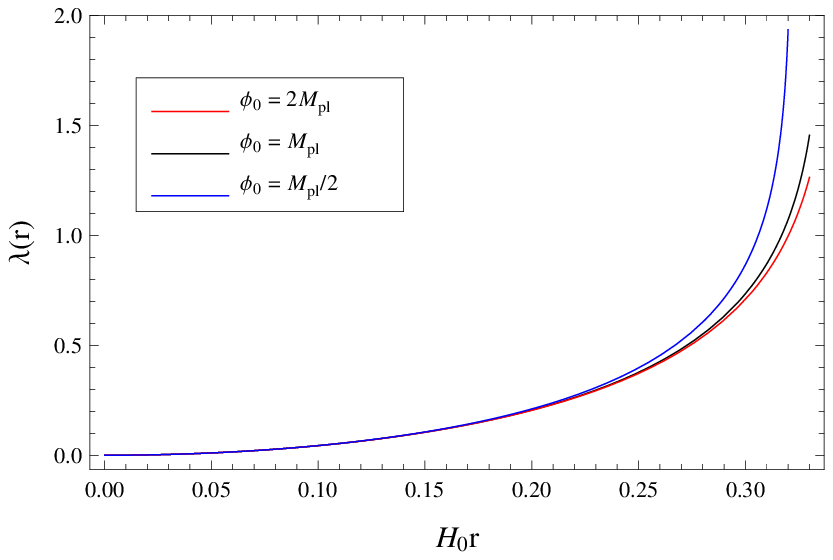}
\end{center}
\end{minipage}
\caption{Examples for the behaviors of $\phi (r)$ (left) and $\lambda (r)$ (right) in the region $r \sim 1/H_0$. 
The constant $M$ is fixed by $M^4=H_0^2M_\mathrm{pl}^2$. The red curve, the black curve, and the blue curve 
represent the cases $\phi _0 = 2 M_\mathrm{pl}$, $\phi _0 = M_\mathrm{pl}$, and $\phi _0 = M_\mathrm{pl}$/2, 
respectively. 
The boundary conditions are applied by using Eqs.~(\ref{q150})-(\ref{q153}) at $r= 10^{-5}H_0^{-1}$ under the assumptions 
$M^4 r^2/(6 \phi _0) \gg \vert c_1 \vert/r , \vert m_1 \vert$ and 
$\kappa ^2 M^8 r^4/\phi _0 ^2 \gg \vert c_2 \vert , \vert r_1 \vert /r, r_0/r$. }
\label{f1}
\end{figure}

In the region that Eq.~(\ref{q150}) is not valid, we need to evaluate the terms 
proportional to $\lambda '$ and $\Phi '$ in Eq.~(\ref{110}). 
However, it would be impossible to solve Eq.~(\ref{110}), analytically. 
Numerical calculations can clarify the behavior of $\phi '(r)$ if we impose the boundary conditions 
by using Eq.~(\ref{q150}). 
The results are shown in Fig~\ref{f1}. Regardless of the value of $\phi _0$, 
$-\phi (r)/M_\mathrm{pl}$ and $\lambda (r)$ go to infinity around $r = 0.3 H_0 ^{-1}$. 
It means that a static metric with a spherical symmetry cannot be a solution in this model around the horizon scale. 
The time evolution of the metric should be taken into account. 
\subsection{Negative power law potential}
In this subsection, we will consider the following form of the potential: 
\begin{equation}
V(\phi)= M^{4+n}\phi ^{-n}, 
\label{q160}
\end{equation}
where $M$ is a constant of mass dimension one and $n$ is a positive number. 
If we assume $V(M_\mathrm{pl})\sim H_0^2 M_\mathrm{pl}^2$, 
which is imposed from the stability of the de-Sitter phase (see \cite{AT}), to realize current accelerated expansion of the Universe, 
we obtain $M \sim 10^{-\frac{46-19n}{4+n}}$GeV. 

The field equation (\ref{q25}) is written by 
\begin{equation}
\phi '' +  \frac{2}{r} \phi ' + n M^{4 +n} \phi ^{-n -1}=0. 
\label{q170}
\end{equation}
This is a nonlinear equation, however, 
we can obtain a particular solution of Eq.~(\ref{q170}) as  
\begin{equation}
\phi (r) = \left [- \frac{2(4+ n)}{n (2+ n)^2 M^{4+n}} \right ]^{-\frac{1}{2+n}} r^\frac{2}{2+n}, 
\label{q200}
\end{equation}
if $n$ is a positive odd number. 
The condition $\vert \phi (r) \vert \ll M_\mathrm{pl}$ can be now rewritten as $r \ll 1/H_0$. 
Substituting Eq.~(\ref{q200}) into Eqs.~(\ref{170}) and (\ref{180}) give the 
following expressions for $\delta \lambda $ and $\delta \Phi$: 
\begin{align}
\delta \lambda (r) = \frac{\kappa ^2 A^2}{2(2+n)} r^\frac{4}{2+n} + C_1, 
\label{q280}\\
\delta \Phi (r) = \frac{4+n}{2n(2+n)(6+n)}\kappa ^2 A^2 r^\frac{4}{2+n} +\frac{r_1}{r} + C_1, 
\label{q290}\\
A \equiv \left [- \frac{2(4+n)}{n(2+n)^2M^{4+n}} \right ]^{-\frac{1}{2+n}}, 
\label{q300}
\end{align}
where $C_1$ and $r_1$ are integration constants. 
The conditions $\vert \delta \lambda \vert , \vert \delta \Phi \vert \ll 1$ yield 
$\vert H_0 r \vert ^\frac{4}{2+n} \ll 1$, $\vert C_1 \vert \ll 1$, and $\vert r_1 \vert \ll r$ if $n=O(1)$. 
Therefore, the solution (\ref{q200}) is valid in the region $\vert r_1 \vert \ll r \ll H_0^{-1}$ 
if $\vert C_1 \vert$ is enough small. 
The parametrized post-Newtonian (PPN) parameter $\gamma$ 
is now expressed as $\gamma =1+(\delta \lambda - \delta \Phi)/(r_0/r)$. 
The tightest constraint for the PPN parameter $\gamma$, $\gamma = 1+(2.1 \pm 2.3)\times 10^{-5}$, 
is imposed by the Cassini experiment \cite{Bertotti:2003rm}. 
In the experiment, the frequency shift of the signal sent by the Cassini spacecraft located at 
8.43 au from the Sun was measured. The signal was transmitted from the Cassini spacecraft to 
the Earth by passing through nearby the Sun. The impact parameter for the Sun is $1.6R_\odot$. 
$\gamma -1$ in this model takes a maximum value at $r=1.6R_\odot$ during the travel of the signal, 
because $\gamma -1$ is proportional to $r^\frac{6+n}{2+n}$ if $\vert r_1/r_0 \vert= \vert r_1 \vert/(GM_\odot)$ is negligible. 
Therefore, $\kappa ^2 A^2 (1.6R_\odot)^{(6+n)/(2+n)}/(GM_\odot) < 10^{-5}$ and $\vert r_1 \vert < 10^{-5}GM_\odot$ would be 
sufficient conditions to pass the constraint.  
By taking into account $\kappa ^2 A^2 \sim H_0^{4/(2+n)}$ and $1.6 R_\odot H_0 \sim 10^{-18}$, 
we obtain $n \lesssim 5$. 
This is a rough estimation, 
however, it is consistent with the concrete calculations of the frequency shift as follows. 
The propagation time of electromagnetic wave from $r=b$ to $r=r_2$ is expressed by 
\begin{equation}
t(r_2,b)= \int ^{r_1}_{b} dr \left [ \frac{\mathrm{e}^{2\lambda (r)-2 \Phi (r)}}
{1-\mathrm{e}^{2 \Phi (r)-2\Phi (b)}\left ( \frac{b}{r} \right )^2 } \right ]^\frac{1}{2} .
\label{time}
\end{equation}
If we express the time (\ref{time}) as $t(r_2,b , \Phi (r), \Psi (r))$, 
the deviation from general relativity $\delta t (r_2, b) \equiv t(r_2,b , \Phi (r), \Psi (r)) 
- t(r_2,b, \Phi _0(r), \Psi _0(r))$, where $\Phi _0(r) = \Psi _0 (r)= -GM_\odot /r$, is written as 
\begin{equation}
\delta t (r_2,b) = \int ^{r_1}_{b} dr \sqrt{\frac{r^2}{r^2-b^2}}\frac{GM_\odot}{r}(\gamma -1). 
\label{timed}
\end{equation}
Substituting the concrete expression of $\gamma$ into Eq.~(\ref{timed}) gives 
\begin{align}
\delta t (r_2,b) = \kappa ^2 A^2 \frac{n^2+5n-4}{2n(2+n)(6+n)} \bigg [ \frac{2+n}{6+n} r_2^{\frac{6+n}{2+n}} 
\bigg ( 1- \frac{1}{2} \frac{\frac{6+n}{2+n}}{2-\frac{6+n}{2+n}}\frac{b^2}{r_2^2} \bigg ) \nonumber \\
+ \sqrt{\pi} b^\frac{6+n}{2+n} \frac{\Gamma \left ( -\frac{1}{2} \frac{6+n}{2+n} \right )}
{2 \Gamma \left ( \frac{1}{2} - \frac{1}{2} \frac{6+n}{2+n} \right )} \bigg ] ,
\end{align}
where we ignored the term proportional to $r_1$. 
Then, the modification for the fractional frequency $\delta \nu / \nu = 2 d \delta t/dt $ \cite{Cassini}, 
which is caused by the motion of the spacecraft and Earth, is approximately expressed as 
\begin{equation}
\frac{\delta \nu}{\nu} \sim \left ( 10^{18-18\frac{6+n}{2+n}} seconds \right ) \times \frac{db}{bdt}, 
\label{fs}
\end{equation}
where $b \simeq 1.6 R_\odot$, $b \ll r_2$, and $n = O(1)$ are assumed. 
The Cassini experiment constrain inside of the parenthesis in Eq.~(\ref{fs}) by $ <10^{-10}$s. 
Therefore, we obtain $n \lesssim 5$. 

\begin{figure}
\begin{minipage}[t]{0.5\columnwidth}
\begin{center}
\includegraphics[clip, width=0.97\columnwidth]{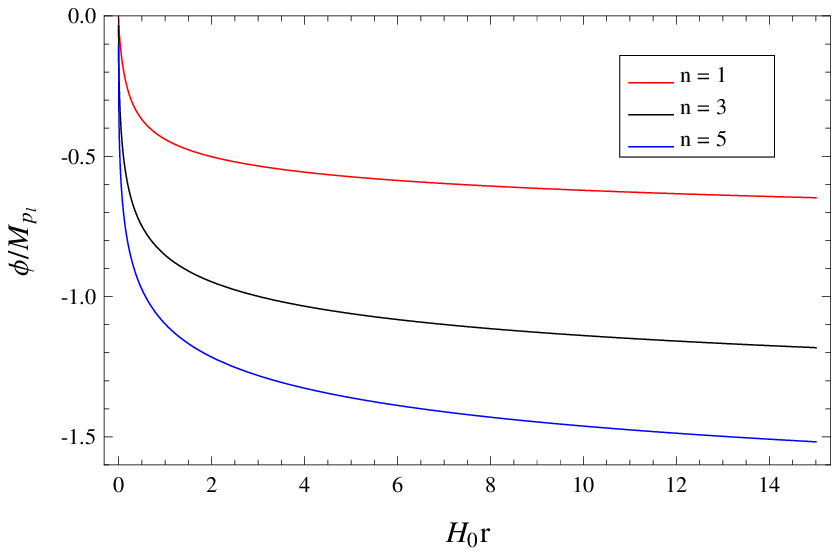}
\end{center}
\end{minipage}%
\begin{minipage}[t]{0.5\columnwidth}
\begin{center}
\includegraphics[clip, width=0.97\columnwidth]{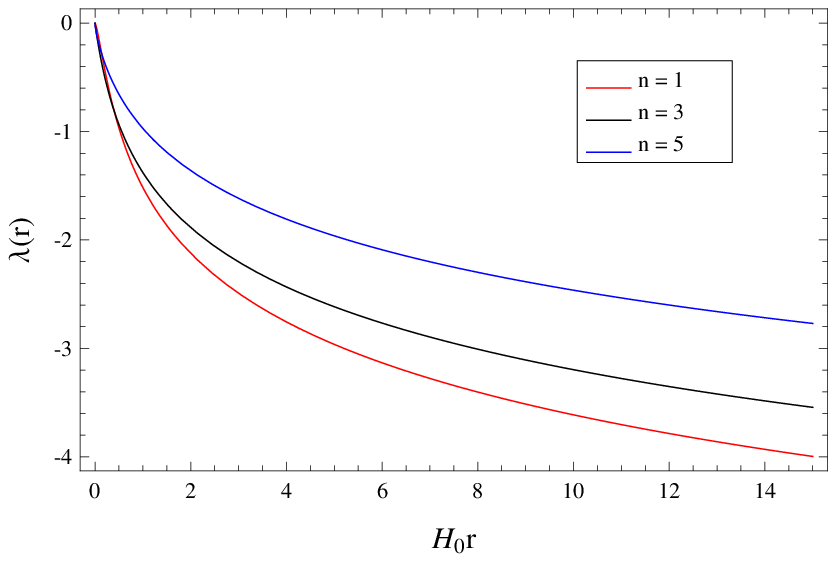}
\end{center}
\end{minipage}
\caption{The behaviors of the scalar field (left) and the metric function $\lambda (r)$ (right) in the case 
$V(\phi)=H_0^2 M_\mathrm{pl}^{2+n} \phi ^{-n}$ in the region $r>H_0^{-1}$. 
The boundary conditions are assigned by Eqs.~(\ref{q200})--(\ref{q300}) at 
$r=10^{-5}H_0^{-1}$. }
\label{f2_1}
\end{figure}

\begin{figure}
\begin{minipage}[t]{0.5\columnwidth}
\begin{center}
\includegraphics[clip, width=0.97\columnwidth]{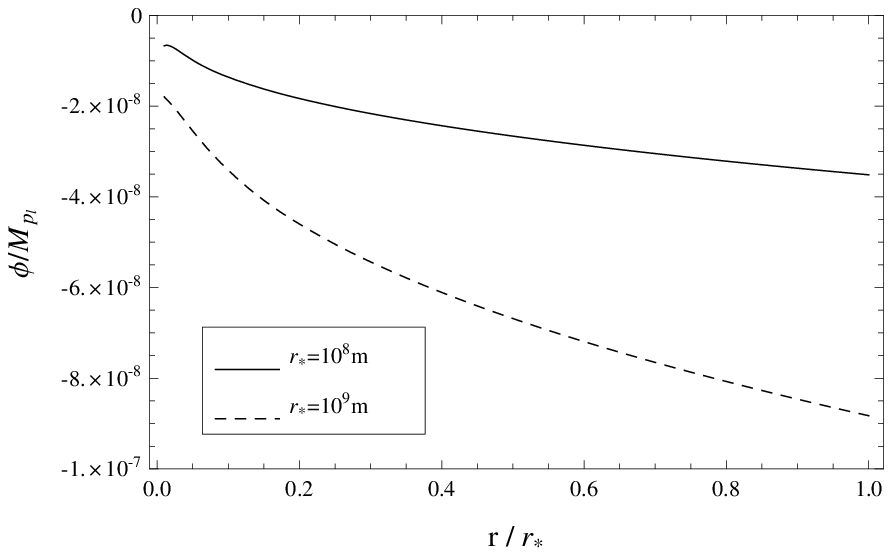}
\end{center}
\end{minipage}%
\begin{minipage}[t]{0.5\columnwidth}
\begin{center}
\includegraphics[clip, width=0.97\columnwidth]{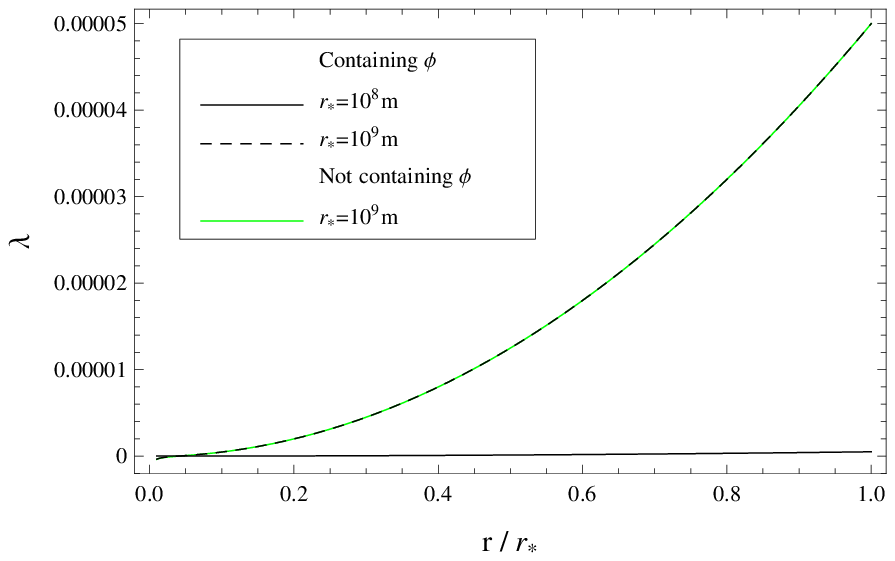}
\end{center}
\end{minipage}
\caption{Interior solutions $\phi (r)$ and $\lambda (r)$ in the case of 
$V(\phi)= H_0^2 M_\mathrm{pl}^{5} \phi ^{-3}$. A constant density 
$\rho _\mathrm{m} (r)= \rho _0 = 1$g/cm$^3$ is assumed for $r<r_*$. The case that there is no scalar field is expressed as 
the green curve in the figure on
the right for comparison. }
\label{f2_2}
\end{figure}

The behavior of the scalar field in the region that Eq.~(\ref{q170}) is not applicable 
is clarified by numerical calculations. 
Figure \ref{f2_1} shows that the $r$ dependence of $\phi (r)$ and $\lambda (r)$ in the region 
$r > H_0^{-1}$. 
The absolute values of $\phi (r)$ and $\lambda (r)$ slowly increase as $r$ becomes large. 
Whereas, Fig.~\ref{f2_2} expresses the behavior of the interior solution when $V(\phi)=M^{4+3}\phi ^{-3}$.  
It shows that $\phi (r)$ and $\phi '(r)$ do not vanish, however, 
the contributions for $\lambda (r)$ from the scalar field are enough suppressed. 
The reason why only the case $n=3$ is depicted is that the energy scale of $\phi (r)$ extremely changes depending on $n$. 
However, the forms of the curves, i.e. $r$ dependence of $\phi (r)$, are not so changed. 
\section{$k$-essence \label{sec4}}
$k$-essence model is a dark energy model expressed by Eq.~(\ref{10}). 
Here, we will shortly consider the case $K(\phi, X)=F(X)$ for simplicity. 
\subsection{Pure kinetic model}
If we choose $K(\phi , X)$ as 
\begin{equation}
K(\phi , X) = -X -s_2 X^2 - s_3 X^3, 
\label{k1}
\end{equation}
where $X= - \mathrm{e}^{-2 \lambda}\phi ^{\prime 2}/2$, 
we obtain the following field equation from Eq.~(\ref{200}):
\begin{align}
\left ( -1+3s_2 \phi ^{\prime 2} -\frac{15}{4}s_3 \phi ^{\prime 4} \right )\phi'' 
+ \frac{2}{r} \left ( -1+s_2 \phi ^{\prime 2} -\frac{3}{4}s_3 \phi ^{\prime 4} \right ) \phi ' =0.
\label{k2}
\end{align}
Equation (\ref{k2}) is simplified as follows when the conditions $\vert X \vert \gg \vert s_2 X^2 \vert \gg \vert s_3 X^3 \vert$ are 
 satisfied: 
\begin{equation}
\phi '' + \frac{2}{r}  \phi ' = 0. 
\label{k3}
\end{equation}
The solution of Eq.~(\ref{k3}) is
\begin{equation}
\phi (r) = \frac{c_1}{r} + \phi _0, 
\label{k4}
\end{equation}
where $c_1$ and $\phi _0$ are arbitrary constants. 
On the other hand, if the conditions $\vert X \vert \ll \vert s_2 X^2 \vert \ll \vert s_3 X^3 \vert$  hold, Eq.~(\ref{k2}) yields 
\begin{equation}
5 s_3 \phi ^{\prime 4} \phi '' + \frac{2}{r} s_3 \phi ^{\prime 5} =0. 
\label{k5}
\end{equation}
The solution of Eq.~(\ref{k5}) is $s_3 \phi ^{\prime 4} =0$ or
\begin{equation}
\phi (r) = b_1 r^\frac{3}{5} + \phi _1, 
\label{k6}
\end{equation}
where $b_1$ and $\phi _1$ are arbitrary constants. 
The expressions of $\delta \lambda (r)$ and $\delta \Phi (r)$ for the
solution (\ref{k4}) are given by 
\begin{equation}
\delta \lambda = \lambda _0 - \frac{c_1^2 \kappa ^2}{4r^2} \left ( 
1-\frac{s_2 c_1^2}{3r^4} +\frac{3s_3 c_1^4}{20r^8} \right ), 
\label{k7}
\end{equation}
\begin{equation}
\delta \Phi = \lambda _0 + \frac{r_1}{r} + \frac{7c_1^4\kappa ^2 s_2}{120} \frac{1}{r^6}-\frac{11c_1^6 \kappa ^2 s_3}{360}
\frac{1}{r^{10}}, 
\label{k8}
\end{equation}
where $\lambda _0$ and $r_1$ are integration constants. 
On the other hand, the solution (\ref{k6}) yields the following expressions 
of $\delta \lambda (r)$ and $\delta \Phi (r)$: 
\begin{equation}
\delta \lambda = - \frac{2187}{50000}\kappa ^2 b_1^6s_3 r^{- \frac{2}{5}} + \lambda _1 
- \frac{81}{500}\kappa ^2 b_1^4 s_2 r^\frac{2}{5} + \frac{3}{20} \kappa ^2 b_1^2 r^\frac{6}{5}, 
\label{k9}
\end{equation}
\begin{equation}
\delta \Phi = \frac{r_2}{r} - \frac{243}{5000}\kappa ^2 s_3 b_1^6 r^{- \frac{2}{5}} 
+ \lambda _1 - \frac{1053}{7000}\kappa ^2 s_2 b_1 ^4 r^\frac{2}{5} + \frac{6}{55}\kappa ^2 b_1^2 r^\frac{6}{5}, 
\label{k10}
\end{equation}
where $\lambda _1$ and $r_2$ are integration constants. Equations (\ref{k7})-(\ref{k10}) show that 
the all of the terms in the expressions of $\delta \lambda (r)$ and $\delta \Phi (r)$ proportional to 
arbitrary constants. Therefore, one may think that the model (\ref{k1}) can easily pass the experimental constraints. 
However, in fact, non-negligible constraints will be imposed by the local gravity constraints. 
First, let us consider Eq.~(\ref{k3}) and the solution (\ref{k4}). 
The conditions $\vert s_2 X \vert , \vert s_3 X/s_2 \vert \ll 1$ yield $\vert c_1^2 s_2 \vert \ll r^4$ 
and $\vert c_1^2 s_3 / s_2 \vert \ll r^4$. 
Whereas, $\vert \delta \lambda \vert \ll 1$ and $\vert \delta \Phi \vert \ll 1$ give $c_1^2 \kappa ^2 \ll r^2$, 
$\vert r_1 \vert \ll r$, and so on. 
Therefore, the applicable region for the solution (\ref{k4}) is 
$\vert r_1 \vert , \vert c_1 \vert \kappa , \vert c_1^2 s_2 \vert ^{1/4}, \vert c_1^2 s_3/s_2 \vert ^{1/4} \ll r$. 
On the other hand, the conditions $\vert s_2 X \vert , \vert s_3 X/s_2 \vert \gg 1$ yield $\vert b_1^2 s_2 \vert r^{-4/5} \gg 1$ 
and $\vert b_1^2 s_3/s_2 \vert r^{-4/5} \gg 1$ if we use the solution (\ref{k6}). 
The conditions for the metric functions $\vert \delta \lambda \vert \ll 1$ and $\vert \delta \Phi \vert \ll 1$, then, 
induce $\kappa ^2 b_1^6 \vert s_3 \vert r^{-2/5} \ll 1$, $\kappa ^2 b_1^4 \vert s_2 \vert r^{2/5} \ll 1$, 
and $\kappa ^2 b_1^2 r^{6/5} \ll 1$. Eliminating $b_1$ from the conditions give 
\begin{equation} 
r \ll \vert s_2 \vert ^{1/2}/ \kappa, \qquad  r \ll \vert s_3/ s_2 \vert ^{1/2}/ \kappa . 
\label{k11}
\end{equation}
Equation (\ref{k11}) is rather a tight constraint for $r$. 
In the usual case, we consider the inner solution (\ref{k6}) as the solution in the solar system. 
Therefore, $\vert s_2 \vert ^{1/2}/ \kappa$ and $\vert s_3/ s_2 \vert ^{1/2}/ \kappa$ should be 
much more than $1$ au. 
If we impose $\vert \delta \lambda (1.6R_\odot) \vert , \vert \delta \Phi (1.6R_\odot) \vert < 
10^{-5} \times GM_\odot/(1.6R_\odot ) \sim 10^{-11}$ as a constraint from the Cassini experiment, we have
\begin{equation} 
1.6 R_\odot \ll 10^{-11} \vert s_2 \vert ^{1/2}/ \kappa, \qquad  1.6 R_\odot \ll 10^{-11} \vert s_3/ s_2 \vert ^{1/2}/ \kappa . 
\label{k12}
\end{equation}
Therefore, the constraints for $s_2$ and $s_3$, 
\begin{equation} 
\frac{1}{(10 \mathrm{eV})^4} \ll \vert s_2 \vert, \qquad \frac{1}{(10 \mathrm{eV})^8} \ll \left \vert s_3 \right \vert ,  
\label{k13}
\end{equation}
are given. The constraints (\ref{k13}) are same as those obtained from the explicit calculations of the frequency shift, 
because most of the contributions for the frequency shift come from $r \sim 1.6 R_\odot$.  
\section{Conclusions \label{sec5}} 
We have considered the behavior of scalar field dark energy with a minimal coupling 
in a static isotropic background. 
In Sec.~\ref{sec2}, we have derived the general equations in $k$-essence model, and have constructed a general 
formalism to investigate the behaviors of the scalar field and the metric functions in the case 
$\rho _\mathrm{m}=p_\mathrm{m}=0 $.  
In Sec.~\ref{sec3}, we have considered quintessence model $K(\phi, X)=-X +V(\phi)$ 
and demonstrated the procedure shown in Sec.~\ref{sec2} in the case $w=const.$ and in the case of negative power law potential. 
The case $w=const.$ is approximately same as the cosmological constant deep inside the horizon scale of the Universe. 
Therefore, it has been shown that this model can easily pass the solar system tests. 
In the case of negative power law potential $V(\phi) \propto \phi ^{-n}$ $(n>0)$, 
we have obtained the particular solutions for the field equation in the region $r \ll 1/H_0$. 
The existence condition for the particular solutions and the constraint from the solar system test 
have shown that the power of $\phi$ should be $-1$, $-3$, or $-5$. 
While, the constraint for $n$ from Hubble parameter mesurement was investigated by 
O. Farooq \textit{et al.} \cite{Farooq:2012ev}. They showed that 
$n \lesssim 1$ at $3 \sigma$ level. Combining our result with their result give a tight constraint on $n$: $n=1$.  
If $n \neq 1$, the model is observationally rejected or does not have a spherically symmetric solution. 

In Sec.~\ref{sec4}, we have shortly considered $k$-essence model 
that only consists of kinetic terms of the scalar field. 
The field equation can be strictly solved in the limit that the higher derivative terms are dominant 
or the lower derivative terms are dominant. 
However, the expansion parameter $s_2$ and $s_3$ are severely constrained by the conditions 
$\vert \delta \lambda \vert , \vert \delta \Phi \vert \ll 1$ and 
$\vert s_2 X \vert , \vert s_3 X/s_2 \vert \gg 1$. 
If we only use either $\vert \delta \lambda \vert , \vert \delta \Phi \vert \ll 1$ or 
$\vert s_2 X \vert , \vert s_3 X/s_2 \vert \gg 1$, there is no constraint on $s_2$ and $s_3$. 
Therefore, evaluating the conditions $\vert \delta \lambda \vert , \vert \delta \Phi \vert \ll 1$ is imperative. 

\section*{Appendix}
The work was supported by the Russian Government Program of Competitive Growth of Kazan Federal University. 

\end{document}